\documentclass[seceqn,secthm,twocolumn,10pt]{baltzer}
\usepackage{graphicx}

\begin{document}
\sloppy

\begin{frontmatter}  

\title{\Large Coexistence of Superconductivity and Antiferromagnetism in Heavy-Fermion Superconductor CeCu$_2$(Si$_{1-x}$Ge$_x$)$_2$ Probed by Cu-NQR \\ --A Test Case for the SO(5) Theory--}

\author[A]{Y.~Kitaoka, K.~Ishida, Y.~Kawasaki}      
\author[B]{O.~Trovarelli, C.~Geibel and F.~Steglich}
\address[A]{Department of Physical Science, Graduate School of Engineering Science, Osaka University, Toyonaka, Osaka, 560-8531, Japan}   
\address[B]{Max-Planck Institute for Chemical Physics of Solids, D-01187 Dresden, Germany}
\runningauthor{}
\runningtitle{}


\begin{abstract}  

We report on the basis of Cu-NQR measurements that superconductivity (SC) and antiferromagnetism (AF) coexist on a microscopic level in CeCu$_2$(Si$_{1-x}$Ge$_x$)$_2$, once a tiny amount of 1\%Ge ($x = 0.01$) is substituted for Si\@. This coexistence arises because Ge substitution expands the unit-cell volume in nearly homogeneous CeCu$_2$Si$_2$ where the SC coexists with {\it slowly fluctuating magnetic waves}. We propose that the underlying exotic phases of SC and AF in either nearly homogeneous or slightly Ge substituted CeCu$_2$Si$_2$ are accountable based on the SO(5) theory that unifies the SC and AF\@.
We suggest that the mechanism of the SC and AF is common in CeCu$_2$Si$_2$\@.

\end{abstract}

\end{frontmatter}



CeCu$_2$Si$_2$ is the first heavy-fermion (HF) superconductor ($T_{\rm c} \sim $ 0.65 K) that was discovered by Steglich {\it et al}.\ in 1979 \cite{Steglich}.
In the antiferromagnetic (AF)-HF CeCu$_2$Ge$_2$ ($T_{\rm N} = 4.15$ K) that has the same lattice and electronic structure as CeCu$_2$Si$_2$, it was found by Jaccard {\it et al.} \cite{Jaccard} that an AF phase touches a superconducting (SC) one at a critical pressure $P_c \sim 7.6$ GPa\@.
Its pressure ($P$)-temperature $(T)$ phase diagram is depicted in Fig.~1a where $T_c \sim 0.65$ K is insensitive in $P_c < P <12$ GPa, but increases at pressures exceeding $P \sim$ 12 GPa\@.
CeCu$_2$Si$_2$ was observed to behave at $P = 0$ much as CeCu$_2$Ge$_2$ does at $P_c$.
Thus the superconductivity in CeCu$_2$Si$_2$ is anticipated to take place close to an AF phase at $P = 0$.
The $P$-$T$ phase diagram in CeCu$_2$Si$_2$ is shown together in Fig.~1a \cite{Thomas,Kitaoka2}, where $T_c$ remains constant in $0 < P < 2$ GPa, followed by a rapid increase in 2 GPa $< P <$ 3 GPa\@.

In recent years, there is increasing evidence that the SC phase is observed even on the border at which the AF order is suppressed by applying the pressure to the AF-HF compounds, CePd$_2$Si$_2$ and CeIn$_3$ \cite{Mathur}.
It was suggested that this kind of SC phase is only viable extremely close to the critical lattice density $D_c$ at which the AF order is suppressed, and in crystals of extremely high purity \cite{Mathur}. 
When a magnetic medium is near $D_c$, the waves of electron-spin density tend to propagate over a long range. 
Thereby, it was argued that the binding of the Cooper pairs could be described in terms of the emission and absorption of waves of electron-spin density. 
However, since these $P$-$T$ phase diagrams were constructed from the resistivity measurement, detailed magnetic nature near $D_c$ or $P_c$ is not experimentally examined yet in these systems.

\begin{figure}
\begin{center}
\includegraphics[width=7.5cm]{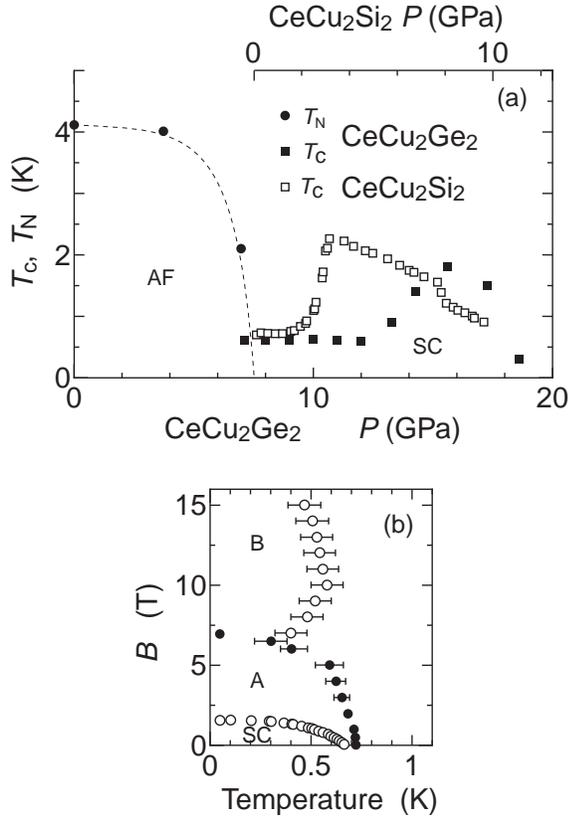}
\caption{(a) Pressure ($P$) versus temperature ($T$) phase diagram in the HF-AF CeCu$_2$Ge$_2$ (ref.~\protect\cite{Jaccard}) and the HF-SC CeCu$_2$Si$_2$ (ref.~\protect\cite{Thomas}).
(b) Magnetic field ($B$) versus $T$ phase diagram \protect\cite{Bruls}.
}
\end{center}
\end{figure}

In either metallic or insulating materials, when approaching an AF phase-transition point ($T_{\rm N}$), $T \rightarrow T_{\rm N}$ or $T_{\rm N} \rightarrow 0$ near $D_c$, {\it slowly fluctuating magnetic waves} become dominant. 
Nuclear spin-lattice relaxation rate, $1/T_1$ can probe their low-frequency components, derived as $1/T_1 \propto \omega_Q/(\omega_Q^2+\omega_{\rm N}^2)$. 
Here $\omega_Q$ and $\omega_{\rm N}$ is a characteristic frequency of magnetic waves and either a nuclear-magnetic-resonance (NMR) or nuclear-quadrupole-resonance (NQR) frequency, respectively.
As $T \rightarrow T_{\rm N}$ or $T \rightarrow 0$ near $D_c$ ($T_{\rm N} \sim 0$), $\omega_Q \propto \sqrt{T-T_{\rm N}}$ and hence $1/T_1$ reveals a sharp peak at $T_{\rm N}$. 
Pronounced reduction of NQR/NMR intensity also becomes precursive because the magnetic waves develop at very low frequencies comparable to $\omega_{\rm N}$. 
Thus the NQR/NMR $T_1$ is one of key measurements to probe the existence of magnetic waves of interest near $D_c$\@. 

Examining characteristics of magnetic medium near $D_c$ is required to prove the existence of magnetically mediated superconductivity.
Extensive experiments including the NQR measurement revealed surprisingly disparate SC and normal-state properties that suggest {\it a breakup} of heavy quasiparticles in nearly homogeneous CeCu$_2$Si$_2$ samples denoted as Type-I and -II as follows \cite{Gegenwart,Ishida2,Kawasaki};
\begin{enumerate}
\item[Type-I:] Polycrystal Ce$_{0.99}$Cu$_{2.02}$Si$_2$(Ce0.99) shows a peak in $1/T_1$ at $T_{\rm c}$, followed by a pronounced deviation from a $1/T_1 \propto T^3$ behavior observed in a Type-II sample.
 The fact that $1/T_1T$ tends toward a constant value well below $T_c$ is indicative of a gapless SC nature with a finite density of states at the Fermi level (see Figs.~3a and 3b).\cite{Ishida2,Kawasaki}. 
At pressures exceeding $P_{A} \sim 0.1$ GPa, the development of {\it slowly fluctuating magnetic waves} is depressed.
Simultaneously, the SC phase that is consistent with a line-node gap model has emerged at zero magnetic field ($B=0$) \cite{Gegenwart,Kawasaki}.
\item[Type-II:] Polycrystal CeCu$_{2.05}$Si$_2$ (Ce1.00) behaves as a best single crystal does, revealing the pronounced reduction in Cu-NQR intensity above $T_c$\@. 
This points to the development of {\it slowly fluctuating magnetic waves} in the normal state \cite{Ishida2,uSR}. 
Below $T_c$ at $B=0$, however, such magnetic fluctuations are expelled \cite{Ishida2,uSR,Bruls}. 
As a result, the typical behavior of $1/T_1 \propto T^3$ is observed in the HF-SC state, consistent with the line-node gap (see Figs.~3) \cite{Ishida2}. 
\end{enumerate}

The NQR experiments do not evidence any static magnetic ordering in the Type-I and -II at least $B$ = 0, but suggest that the former is extremely close to the border of a magnetic ordering. 
When the magnetic field exceeds an upper critical field $H_{\rm c2}$ for the Type-I and -II at $P = 0$ to suppress the superconductivity, on the other hand, the measurements of elastic constant, thermal expansion \cite{Bruls}, $1/T_1T$ \cite{Kitaoka2} and the specific heat \cite{Kobayashi} revealed an evolution from a SC phase into some magnetic phase that was marked as "phase A". 
This phase emerges below $T_A$ close to the $T_c$ at $B = 0$. 
Furthermore, the resistivity measurement on a single crystal suggested that "phase A" seems to behave as a spin-density-wave (SDW)-type ordering \cite{Gegenwart}, and phase B is marked to be also magnetic in origin \cite{Gegenwart,Bruls}. 
These field-induced phases are presented in Fig.~1b\@.

Applying the pressure compresses the unit-cell volume, $V$ and increases the lattice density $D = 1/V$. 
By contrast, partial Ge substitution for Si expands linearly $V$ and hence decreases $D$ \cite{Trovarelli}. 
The magnetic and SC phase diagram in CeCu$_2$(Si$_{1-x}$Ge$_x$)$_2$ in $0.02 < x <1$ was reported using various type of measurements. 
Coexistence of "phase A" and the superconductivity was suggested in a microscopic scale \cite{Trovarelli}.

In order to gain further insight into these exotic SC and magnetic phases and to clarify characters of magnetic fluctuations near $D_c$, we have investigated CeCu$_2$(Si$_{1-x}$Ge$_x$)$_2$ through the Cu-$T_1$ and NQR-spectrum measurements. 
We show that the SC and AF order coexist on a microscopic level, once a tiny amount of 1\%Ge ($x = 0.01$) is substituted for Si\@. 
We propose that these exotic phases found in CeCu$_2$Si$_2$, that have been long standing issues over a decade, are accountable based on the SO(5) theory that unifies antiferromagnetism and d-wave superconductivity \cite{Zhang}. 
We shed new light on SC states in strongly-correlated-electron systems near $D_c$\@.
%
\begin{figure}
\begin{center}
\includegraphics[width=7.2cm]{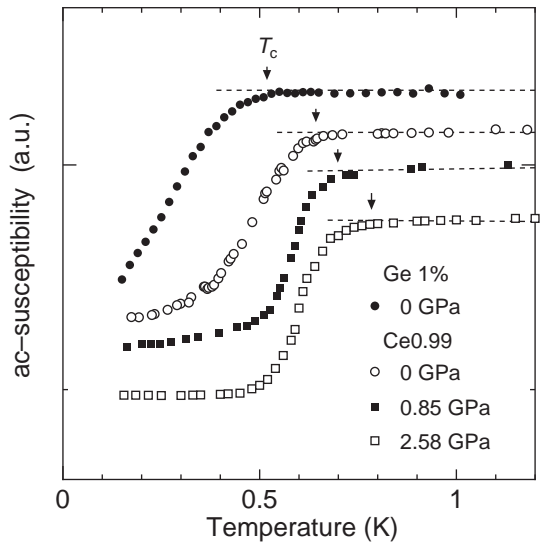}
\caption{Representative ac-susceptibility for 1\%Ge-doped CeCu$_2$Si$_2$ along with those for the Type-I Ce0.99 at ambient pressure ($P = 0$), $P = 0.85$ and 2.58 GPa \protect\cite{Kawasaki}. 
Note that the SC natures in the Type-I Ce0.99 in $0.85$ GPa $< P$ are typical for other HF-SC compounds with the line-node gap reported thus far \protect\cite{Gegenwart,Kawasaki}.}
\end{center}
\end{figure}

We here report Cu-NQR study on CeCu$_{2}$(Si$_{1-x}$Ge$_x$)$_2$ at $x$ = 0.01, 0.02, 0.1 and 0.2 which are the same as used in the previous work \cite{Trovarelli}. 
High-frequency ac-susceptibility (ac-$\chi$) was measured by using an in-situ NQR coil to ensure a bulk nature of the superconductivity. 
The bulk SC nature at $x$ = 0.01 and 0.02 is corroborated by a comparable size of SC diamagnetism at low temperatures among the ac-$\chi$ data at various values of pressure on the Type-I Ce0.99 as presented in Fig.~2 where the data at $x$ = 0.01 are represented. 
Respective SC transition temperature was determined to be 0.5 and 0.4 K for $x$ = 0.01 and 0.02. 
Note that the SC natures in the Type-I Ce0.99 in $0.85$GPa $< P$ are typical for other HF-SC compounds with the line-node gap \cite{Gegenwart,Kawasaki}. 
$T_{\rm c} = $ 0.2 K and 0.15 K was reported for $x$ = 0.1 and 0.15 in ref.~\cite{Trovarelli}, respectively. 
The samples were moderately crushed into grains with diameters larger than 100 $\mu$m in order to avoid some crystal distortion. 
The Cu-NQR spectrum was obtained by plotting spin-echo intensity as a function of frequency at $B = 0$ and in a $T$ range 0.1 -- 4.2 K\@. 
$T_1$ was measured by the conventional saturation-recovery method at $B$ = 0 in $T = $ 0.05 -- 50 K\@.

\begin{figure}
\begin{center}
\includegraphics[width=7.1cm]{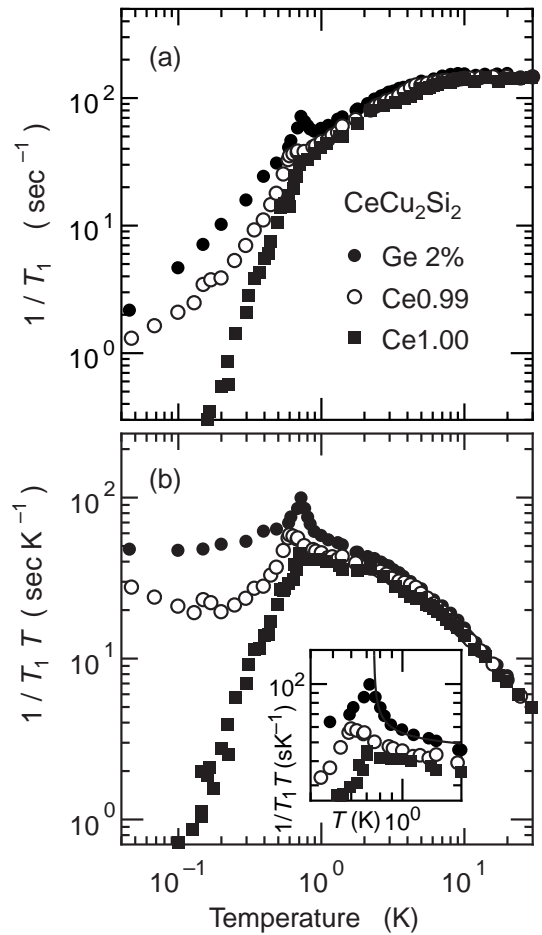}
\caption{Temperature ($T$) dependence of (a) $(1/T_1)$ and (b) $(1/T_1T)$ for 2\%Ge-doped CeCu$_2$Si$_2$ ($T_{\rm N} = 0.75$ K and $T_{\rm c} = 0.4$ K), Ce$_{0.99}$Cu$_{2.02}$Si$_2$ ($T_{\rm c} = 0.65$ K), and CeCu$_{2.05}$Si$_2$ ($T_{\rm c} = 0.7$ K) \protect\cite{Ishida2}.
The inset indicates the $1/T_1T$ vs $T$ plot in both expanded scales.
Solid line indicates a fit by $1/T_1T = 6.6/\protect\sqrt{T-0.75}+43$ that is consistent with the data in $T_N < T < 2$ K for $x = 0.02$.
}
\end{center}
\end{figure}

\begin{figure}
\begin{center}
\includegraphics[width=7.5cm]{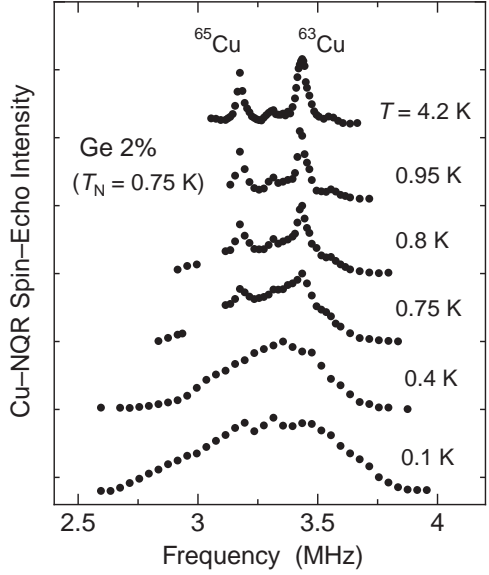}
\caption{$T$ dependence of Cu-NQR spectrum. 
The spectrum shows the hyperfine broadening due to the emergence of magnetic ordering below $T_N \sim$ 0.75 K\@.
}
\end{center}
\end{figure}

Figs.~3a and 3b show the respective $T$ dependence of $1/T_1$ and $1/T_1T$ for $x = 0.02$ along with those for Ce0.99 and Ce1.00 \cite{Ishida2}. 
Single $T_1$ value was determined from the simple exponential recovery of nuclear magnetization in a measured $T$ range except below $T_N$. 
Clear peaks in both quantities for $x$ = 0.02 ($x$ = 0.01, not shown) emerge at $T_{\rm N}$ = 0.75 (0.65) K, probing the development of slow magnetic fluctuations towards an AF-type phase-transition point. 
The data for $x$ = 0.02 in the normal state ($T_N < T < 2$ K) appear to be consistent with a behavior $1/T_1T = 6.6/\sqrt{T-0.75}+43$ (solid line in the inset of Fig.3b). 
A nearly or weakly AF itinerant-electron model (SCR theory) was applied for the understanding of the unusual normal and SC properties of the HF compounds nearby an AF instability \cite{Moriya,Nakamura}. 
In this SCR scheme, $1/T_1T \propto\sqrt{\chi_Q(T)}\propto 1/\sqrt{T-T_N}$ was predicted, since $\chi_Q(T)$ follows a $C/(T-T_N)$. 
In fact, such a behavior was reported in the AF-HF superconductor UNi$_2$Al$_3$ \cite{Kyogaku}. 
The fact that the $1/T_1T$ data for the Ge doped sample are fit by including the term $6.6/\sqrt{T-0.75}$ demonstrates a long-range nature of its magnetic ordering. 
Furthermore, an emergence of the AF-type order below $T_{\rm N}\sim 0.75$ K is corroborated by the appearance of internal fields ($H_{int}$) that are probed by the significant increase of NQR spectral width. 
These spectra are indicated in Fig.~4\@. 
It is at present hard to identify its magnetic structure from the hyperfine-broadened shape in the Cu-NQR spectrum below $T_N$, although some SDW-like ordering is likely from a distribution in $H_{int}$. 

\begin{figure}
\begin{center}
\includegraphics[width=7.0cm]{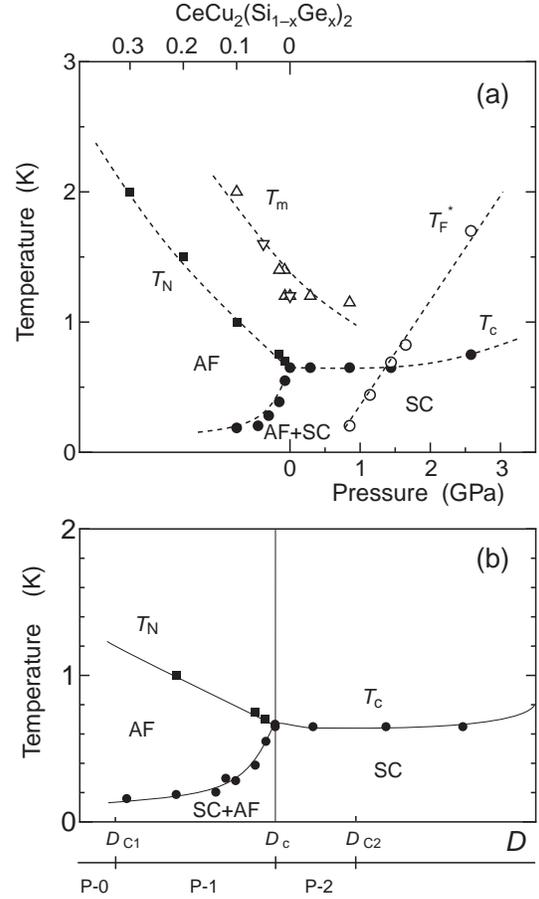}
\caption{(a) The AF- and SC-phase diagrams in CeCu$_2$(Si$_{1-x}$Ge$_x$)$_2$ and in the Type-I sample under the pressure. 
Also shown are an effective Fermi temperature $T_F^*$ below which a $T_1T$ = const behavior would be expected and $T_m$ below which slowly fluctuating magnetic waves are dominant \protect\cite{Kawasaki}. 
(b) The AF- and SC-phase diagram as the function of lattice density $D$ in CeCu$_2$(Si$_{1-x}$Ge$_x$)$_2$ ($D < D_c$) and in CeCu$_2$Si$_2$ ($D_c \leq D$) under $P$. Note $D \propto 1/V$ where $V$ is the unit-cell volume and $D = D_{\rm Si}[1-(V_{\rm Ge}-V_{\rm Si})x/V_{\rm Ge}]$ in the former.}
\end{center}
\end{figure}

Thus determined $T_{\rm N}$ is plotted in Fig.~5a where the phase diagram of CeCu$_2$(Si$_{1-x}$Ge$_x$)$_2$ is indicated together with that under $P$ \cite{Kawasaki}. In the figure are shown an effective Fermi temperature $T_F^*$ below which a $T_1T$=const behavior would be expected and $T_m$ below which slowly fluctuating magnetic waves are dominant \cite{Kawasaki}. 
We conclude that $T_{\rm c} < T_{\rm N}$ and hence the AF and SC order coexist, once Ge is slightly substituted for Si in the Type-I Ce0.99 and eventually its $D$ becomes smaller than $D_c$ ($D < D_c$).
We should also remark that the $T_1T$ = const behavior in the SC state (see Fig.3b) evidences a gapless SC state with a finite density of states \cite{coexistence}. 
This contrasts with the line-node gap state that is observed in the Type-I Ce0.99 under 0.85 GPa $< P$ and the Type-II Ce1.00 (see Figs.~3). 
The present NQR results have, for the first time, given convincing experimental ground that the SC and AF order coexist in $D < D_c$ in {\it a microscopic scale}.
It is noteworthy that a temperature $T_A$ below which "phase A" emerges \cite{Gegenwart,Bruls,Trovarelli} is higher than either $T_N$ or $T_c$, but comparable to $T_m$\@. 
In this context, "phase A" that was proposed to exist even at $B = 0$ from the various bulk measurements \cite{Gegenwart,Trovarelli} is shown to be neither an AF- nor SDW-type phase at $B$ = 0 \cite{Ishida2,Kawasaki}. 

Noting that $D = D_{\rm Si}[1-(V_{\rm Ge}-V_{\rm Si})x/V_{\rm Ge}]$ for the Ge doping and $D$ increases with the pressure, we present a combined phase diagram as the function of $D$ in Fig.~5b\@. 
In the phase diagram of Fig.~5b, we denote the AF-type phase in $D < D_{\rm c1}$ as (P-0), the coexisting phase of SC and AF order in $D_{\rm c1} < D < D_c$ as (P-1), and the SC phase at $B = 0$ in $D_c \leq D \leq D_{\rm c2}$ as (P-2). 
In (P-2), note that the SC state evolves into "phase A" and subsequent "phase B" as the magnetic field increases exceeding $H_{\rm c2}$ as seen in Fig.~1b \cite{Bruls}. 
The SC region is widespread in $D_{c1} < D$, whereas the AF order disappears at $B = 0$ in $D_c < D$. 
Thus with increasing $D$, $D_{c1}$ and $D_{c}$ is characterized as the respective critical lattice density at which the superconductivity sets in and the AF order is suppressed at $B = 0$. 
We suppose that $D_{c2}$ is the third critical density exceeding which a field induced "phase A" no longer appears \cite{Gegenwart}.\\[3mm]
%

{\it Phenomenology on the SO(5) theory }

We now propose that the SO(5) theory that was proposed by Shou-Cheng Zhang \cite{Zhang} is promising in understanding the underlying phases found in CeCu$_2$Si$_2$\@. 
In this unified theory of AF and SC order, a concept of {\it superspin} is introduced. 
It is a five-dimensional vector, $\vec n_i = (n_1, n_2, n_3, n_4, n_5)$. 
Magnitude of the vector is preserved with the constraint: $\sum_{i=1}^5 n_i ^2=1$ or $|\Delta|^2+|S_Q|^2 = 1$.
Here $|\Delta|^2 = n_1^2+n_5^2$ and $|S_Q|^2 = n_2^2+n_3^2+n_4^2$ is the respective amplitude of the SC and AF order parameter (OP). 
{\it In this new theory, the AF and SC order are complementary and the same magnetic interaction leading to an AF state gives rise to pair binding.} In the low-energy asymmetric SO(5) Hamiltonian, a $\mu$ term, $-2\mu Q_c$ (where $Q_c$ is the total charge) is added, appearing as a "gauge coupling'' in the field theory.
Then an effective potential energy is given by $V_{\rm eff} = -\frac{g}{2}|S_Q|^2-\frac{(2\mu)^2}{2}|\Delta|^2(\chi_c|\Delta|^2 + \chi_{\pi}|S_Q|^2)$. Here $\chi_c$ is the charge compressibility and $Q_c= 2\mu(D)\chi_c$ as the function of lattice density $D$\@.
$\chi_\pi$ is the newly introduced "$\pi$'' susceptibility which enables to rotate the AF-OP into SC-OP and vice versa. 
In order for the SO(5) symmetry to be approximately valid, these parameters are close in value. 
For $|\Delta|^2 = 0$, since the AF phase is selected, a coupling constant $g$ in the presence of explicit symmetry breaking is fixed to $g > 0$. 
Zhang predicted that the richness of the phase diagram at $T = 0$ comes entirely from varying $\mu(D)$ in the present case as follows; (1) [SO(5)-0] : an AF phase in $[2\mu(D)]^2 \leq g/\chi_{\pi}$, (2) [SO(5)-1] : a coexisting phase of AF and SC order in $g/\chi_{\pi} < [2\mu(D)]^2 < g/(2\chi_c-\chi_{\pi})$ when $\chi_c <\chi_{\pi} < 2\chi_c$, and (3) [SO(5)-2] : a SC state in $(g/\chi_c) \leq [2\mu(D)]^2$ when $\chi_\pi \leq \chi_c$.

\begin{figure}[htbp]
\begin{center}
\includegraphics[width=7.6cm]{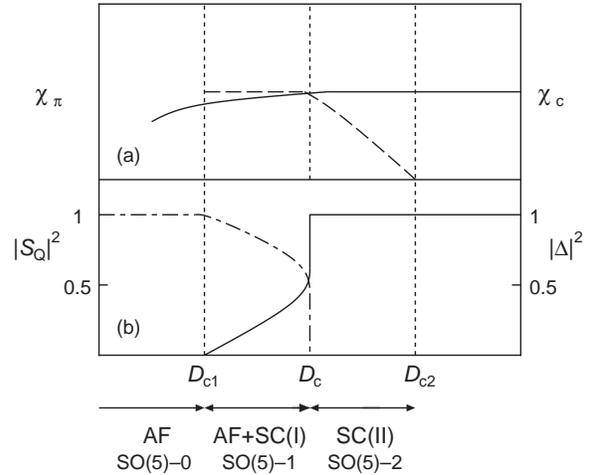}
\caption[]{(a) Schematic variation in the charge compressibility $\chi_c$ (solid line) and the ``$\pi$'' susceptibility $\chi_{\pi}$ (dash line) (see text) as the function of $D$ to be consistent with the complex phases (P-0), (P-1) and (P-2) in Fig. 5b\@. 
Phases predicted at $T = 0$ on the SO(5) theory are assigned as follows; [SO(5)-0] to the AF phase in (P-0), [SO(5)-1] to the coexisting phase of AF and SC order in (P-1), and [S0(5)-2] to the SC phase in (P-2). (b) Possible variations in the AF-OP, $|S_Q|^2$ (dash-dot line) and the SC-OP, $|\Delta|^2$ (solid line) against $D$ in [SO(5)-1].
}
\end{center}
\end{figure}

We point out that the underlying phases observed in CeCu$_2$Si$_2$ (Fig.~5) are accountable on the SO(5) theory by assuming $\chi_{\pi}(D_{c1}) = g/[2\mu(D_{c1})]^2$ at $D_{\rm c1}$, $\chi_{\pi}(D_c)=\chi_c(D_c)=g/[2\mu(D_c)]^2 \equiv \chi_0$ at $D_c$ and $\chi_{\pi}=0$ at $D_{\rm c2}$. 
A point is that $\chi_c(D)$ and $\chi_\pi(D)$ are assumed to vary as the function of $D$ as schematically indicated in Fig.~6a. 
Here note that the variation in $\chi_c$ might be moderate because of metallic states in both sides of $D_c$ and $\chi_{\pi}(D_{\rm c1}) \sim [1+2(\frac{\partial\mu}{\partial D})\frac{D_c-D_{\rm c1}}{\mu(D_c)}]\chi_0$ is experimentally determined. 
Indeed, $\chi_{\pi}(D_{c1})$ and $\chi_0(D_c)$ are close in values because $(D_c-D_{c1})/D_c\sim 0.01$ and $(\frac{\partial\mu}{\partial D})$ is very small \cite{Kawasaki,Trovarelli}. 
The SO(5) symmetry is hence supposed to be approximately valid.
We assign the respective phase (P-0), (P-1) and (P-2) in Fig.~5b to [SO(5)-0], [SO(5)-1] and [SO(5)-2]\@. 
Schrieffer, Wen and Zhang described the phase [SO(5)-1] as a spin-bag phase in terms of pairing the eigenstates of the AF background \cite{Schrieffer}. 
We suggest that the AF-OP and SC-OP change as the function of $D$ in the phase [SO(5)-1] as indicated in Fig.~6b\@.

Based on the SO(5) theory, it is possible to account for the exotic SC phase in the Type-I Ce0.99 that exhibits anomalies associated with {\it slowly fluctuating magnetic waves} \cite{Ishida2,uSR}. 
In the case that $\chi_{0}\simeq\chi_c(D)\leq\chi_{\pi}(D)$ and $\mu(D) \simeq \mu(D_c)$, we may indicate that the AF and SC order coexist with $|\Delta|^2 = |S_Q|^2 = 1/2$ at $T = 0$. 
For a finite temperature in $T_{\rm c} < T \leq T_{\rm N}$, on the other hand, we suggest that SC fluctuations prevent an onset of AF order when its characteristic frequencies of fluctuations are imposed so that $\omega_Q(T) \simeq \omega_{\rm SC}(T) < [2\mu(D_0)] \sqrt{\frac{(T_{\rm N}-T)}{T_{\rm N}}|\Delta(T)|^2\cdot[|\Delta(T)|^2+\frac{\chi_{\pi}}{\chi_0}|S_Q(T)|^2]}$. Here note $|S_Q(T)|^2 < |\Delta(T)|^2 < 1$. 
Therefore, {\it slowly fluctuating magnetic waves} may survive even below $T_c$ as was actually observed in the Type-I Ce0.99 \cite{Ishida2,Kawasaki,uSR}. 
In the case that $\chi_c(D)<\chi_{\pi}(D)$ and $\mu(D)<\mu(D_c)$, however, the onset of AF order becomes energetically more stable and coexists with the SC phase in $T < T_c < T_N$\@. 
The SO(5) theory accounts for the unusual SC and magnetic phenomena in the Type-I Ce0.99 \cite{Gegenwart,Ishida2,uSR} as well as the coexistence of SC and AF order in CeCu$_2$(Si$_{x-1}$Ge$_x$)$_2$\@. 
In this context, the Type-I Ce0.99 is anticipated to be quite close to the condition for the SO(5) symmetry to be exact, i.e. $\chi_{\pi}=\chi_c$.

We next argue based on the SO(5) theory \cite{Hu} the magnetic-field induced "phase A" in the Type-II samples where $D_c \leq D$\@. 
In the {\it superspin} picture, starting from a SC state at $B_z$ = 0, $\vec n$ lies in the ($n_1$, $n_5$) SC-subplane. 
As $B_z$ increases, a uniform-spin polarization is induced along a SC vortex over a London penetration depth, $\lambda$. 
At the same time, the staggered polarization, $n_2^2+n_3^2 = |S_Q^{\perp}|^2$ perpendicular to $B_z$ is induced over a SC coherence length, $\xi$ where $|\Delta(r)|^2 = |\Delta|^2[1-\exp{(-r/\xi)}]$. 
This is because the magnitude of the {\it superspin} is preserved. 
In the presence of $B_z$ and $\mu$, by taking a general Ginzburg-Landau potential of an approximate SO(5) model \cite{Hu}, it was argued that the magnetic-field induces either a first-order-phase transition from the SC state to the AF state at a critical value of the magnetic-field $B_c$ or two second-order-phase transitions with an intervening mixed phase region where the SC and AF order coexist.
 It is here proposed that "phase A" in the Type-II (see Fig.~1b) corresponds to this magnetic-field induced AF state in the [SO(5)-2], although further investigations are required to determine a magnetic structure of "phase A". 
Indeed, we believe that the phenomenological approach based on the SO(5) theory has given a coherent interpretation for (1) the coexisting phase of SC and AF order in CeCu$_2$(Si$_{1-x}$Ge$_{x}$)$_2$, (2) the exotic SC phase in the Type-I Ce0.99 near $D_c$ where we propose $\chi_{\pi}\sim\chi_c$, and (3) the magnetic-field induced "phase A" in the Type-II Ce1.00.\\[3mm]
{\it Implication of new concept for superconductivity}

Unconventional interplay between superconductivity and magnetism found in the homogeneous CeCu$_2$Si$_2$ has been a long standing issue to resolve over a decade. 
This was shown to originate from that the system is just on the border of the AF phase ($D \sim D_c$) and the SC and AF coexist once $D < D_c$. 
We have proposed that the SO(5) theory constructed on the quantum-field theory (refs.~\cite{Zhang,Hu,Zacher}) is applicable to give a coherent interpretation for these exotic phases found in the strongly correlated HF-superconductor CeCu$_2$Si$_2$\@. 
In this context, we may suggest that the superconductivity in CeCu$_2$Si$_2$ could be mediated by the same magnetic interaction leading to the AF state in CeCu$_2$(Si$_{1-x}$Ge$_{x}$)$_2$\@. 
This is in marked contrast to the BCS one, in which the pair binding is mediated by phonons, vibration of lattice density. 
We believe that this model could shed further light on the current models of magnetic mechanism for the superconductivity in other strongly-correlated-electron systems, involving even the high-$T_c$ cuprates \cite{Zhang,Zacher} and the organic \cite{Murakami} superconductors.

We thank H.~Kohno, K.~Miyake, N.~Nagaosa and S.~Murakami for stimulating discussions and valuable comments. 
This work was supported by the COE Research (10CE2004) in Grant-in-Aid for Scientific Research from the Ministry of Education, Sport, Science and Culture of Japan.


%

\begin{thebibliography}{00}
\bibitem{coexistence}
{\footnotesize We note that the absence of any clear reductions in $1/T_1$ and $1/T_1T$ below $T_{\rm c} \sim 0.4$ K ($x = 0.02$) is in 
remarkable contrast to the bulk-SC signatures.
This is because low-frequency magnetic excitations in the AF state that govern the nuclear relaxation channel persist even in the SC state. 
This ensures that the SC phase is not spatially segregated, but coexists with the AF order on a microscopic scale.}
\end{thebibliography}
\end{document}